\documentclass[useAMS,usenatbib]{mn2e}

\usepackage{times,graphicx,amssymb,natbib,amsmath}
\pdfoutput = 1

\newcommand{\degrees}{^{\circ}}
\newcommand{\msol}{M_{\rm \odot}}
\newcommand{\mjup}{M_{\rm Jup}}

\newcommand{\mearth}{M_{\rm \oplus}}
\newcommand{\cotwo}{\rm CO_2}

\title[Advanced Climate Models of Earth-like Exomoons]{Exomoon Climate Models with the Carbonate-Silicate Cycle and Viscoelastic Tidal Heating}
\author[Duncan Forgan and Vera Dobos]{Duncan Forgan $^{1}$\thanks{E-mail:dhf3@st-andrews.ac.uk} and Vera Dobos$^{2}$  \\
$^{1}$Scottish Universities Physics Alliance (SUPA), School of Physics and Astronomy, University of St Andrews, North Haugh, KY16 9SS, UK \\
$^{2}$Konkoly Thege Miklos Astronomical Institute, Research Centre of Astronomy and Earth Sciences, Hungarian Academy of Sciences, \\
H–1121 Konkoly Thege Mikl\'os \'ut 15-17, Budapest, Hungary \\
}

\begin{document}

\date{Accepted}

\pagerange{\pageref{firstpage}--\pageref{lastpage}} \pubyear{}

\maketitle

\label{firstpage}

\begin{abstract}

\noindent The habitable zone for exomoons with Earth-like properties is a non-trivial manifold, compared to that of Earth-like exoplanets.  The presence of tidal heating, eclipses and planetary illumination in the exomoon energy budget combine to produce both circumstellar and circumplanetary habitable regions.  Analytical calculations suggest that the circumplanetary habitable region is defined only by an inner edge (with its outer limits determined by orbital stability).  Subsequent calculations using 1D latitudinal climate models indicated that the combined effect of eclipses and ice-albedo feedback can produce an outer edge to the circumplanetary habitable zone.  But is this outer edge real, or an artefact of the climate model's relative simplicity?

We present an upgraded 1D climate model of Earth-like exomoon climates, containing the carbonate-silicate cycle and viscoelastic tidal heating.  We conduct parameter surveys of both the circumstellar and circumplanetary habitable zones, and we find that the outer circumplanetary habitable edge remains provided the moon's orbit is not inclined relative to that of the planet.  Adding the carbonate-silicate cycle pushes the circumplanetary habitable zone outward, by allowing increases in atmospheric partial pressure of carbon dioxide to boost the greenhouse effect.  Viscoelastic tidal heating widens the habitable zone compared to standard, fixed-Q models.  Weakening the tidal heating effect due to melting allows moons to be habitable at higher eccentricity, and pushes the inner circumstellar and circumplanetary habitable zone boundary inward.

\end{abstract}


\section{Introduction}

\noindent At the time of writing, extrasolar moons (exomoons) remain undetected.  A variety of detection methods for exomoons exist, including transit timing and duration variations \citep{Simon2007,Kipping2009,Heller2014}, microlensing \citep{Liebig2010}, and direct imaging \citep{Peters2013, Agol2015}.  Efforts from several teams \citep{Kipping2015, Hippke2015} have produced only upper limits on the occurrence of relatively massive moons (see e.g. \citealt{Weidner2010}).  A tentative detection of an exomoon orbiting a free floating planet via microlensing cannot be confirmed due to uncertainties in the system's distance from Earth \citep{Bennett2013}.

However, these detection methods are continuing to sweep further into exomoon parameter space, in a manner quite analogous to early exoplanet detection efforts, which eventually yielded the pulsar planets around PSR 1257+12 \citep{Wolszczan1992} and the hot Jupiter 51 Pegasi b \citep{Mayor1995}.  Moons are relatively common in our own Solar System, and they are predicted to be a common outcome of the planet formation process, their formation in circumplanetary discs being something of a microcosm of planet formation in circumstellar discs (e.g. \citealt{Mosqueira2003, Mosqueira2003a,Ward2010,Canup2006}).

Therefore, in anticipation of those tantalising first detections of exomoons, we can consider the possibility that these objects could be habitable.  The potential for liquid water in Solar system moons such as Europa \citep{Melosh2004}, Enceladus \citep{Iess2014,Thomas2015a}  and Ganymede \citep{Saur2015} are the consequence of tidal heating induced by their parent planets, and such subsurface habitats may be extremely common in the Universe \citep{Scharf2006}.  Relatively massive exomoons could host a substantial atmosphere like the Earth, and could host a similar biosphere.

The habitable zone (HZ) for such moons would have a qualitatively different structure to that for planets.  The standard planetary HZ concept relies principally on the received stellar flux, the planet's atmospheric albedo and its emission of longwave radiation.  1D altitudinal calculations using the latest line radiative transfer algorithms and atmospheric data allow the construction of habitable zones that depend only on the host star's luminosity and effective temperature, with the other parameters assumed to be Earth-like (\citealt{Kopparapu2013} although some recent work by \citealt{Kopparapu2014} has extended this to more and less massive planets).

The habitable zone for moons is significantly affected by the properties of the host planet, as well as the properties of the host star.  As mentioned above, tidal heating can create oases of habitability beyond the edge of the planetary habitable zone \citep{Reynolds1987,Scharf2006}, and roast moons inside the planetary habitable zone (e.g. \citealt{Heller2013b}) .

If the moon's orbital inclination relative to the planetary orbit is low, then eclipses can be relatively frequent, reducing the orbit averaged stellar flux by a few percent \citep{Heller2012}.  Moons on very high inclination orbits (orbiting retrograde) are typically warmer than low inclination, prograde moons, as retrograde moons experience shorter, more frequent eclipses \citep{Forgan_moon1}.  The planet itself is likely to emit strongly in the infrared through a combination of reradiated and reflected starlight, pushing the habitable zone further from the star \citep{Heller2013}.

Combining these factors requires us to consider the circumstellar habitable zone \emph{and} the circumplanetary habitable zone.  From analytical calculations of the total flux received from all sources, \citet{Heller2013b} showed that the circumplanetary habitable region only has an inner edge where tidal heating induces a runaway greenhouse effect.  The habitable region extends to planetocentric distances of order half a Hill Radius \citep{Domingos2006} where the moon can no longer execute stable orbits around the planet.  The circumplanetary habitable zone consists of only an inner edge, with the outer edge as distant as dynamically permissible.

\citet{Forgan2014a} showed using 1D latitudinal energy balance climate modelling that this picture is correct if the planet resides well inside the circumstellar habitable zone.  If the planet orbits near the outer edge of the circumstellar habitable zone \emph{and the orbits of the planet and moon are close to coplanar}, the combination of eclipses and the climate's ice-albedo feedback mechanism can induce a snowball state in moons from which they struggle to escape.  This defines an outer circumplanetary edge that exists for a specific range of planetary and lunar orbital parameters, in particular the lunar inclination.  It should be noted that in general a single eclipse was insufficient to generate a snowball state, but rather the cumulative effect of many eclipses, with each eclipse slightly increasing the amount of frozen surface after the moon returns to orbital longitudes that possess stellar illumination.

However, given that this edge exists primarily due to a strong positive feedback mechanism in the climate model, we must be cautious of the model's veracity.  In particular, we must take care to include all feedback mechanisms, both positive and negative.  The ice-albedo feedback mechanism is a strong positive mechanism which encourages rapid freezing of a planet's surface as increasing ice cover increases the local albedo.

The carbonate-silicate cycle is a negative feedback mechanism that regulates the atmospheric partial pressure of carbon dioxide ($\cotwo$).  $\cotwo$ is removed from the atmosphere through precipitation and silicate weathering, which occurs at increasing rates with increasing temperature.  This weathering process returns the carbon to the oceans, where it forms carbonates on the sea floor and is subsequently subducted into the mantle.  The cycle is completed by volcanism expelling $\cotwo$ into the atmosphere.  This cycle adjusts the $\cotwo$ partial pressure in response to changes in temperature, acting as a ``thermostat'' to damp these fluctuations.  It is immediately clear then that $\cotwo$ adjustment could be a means by which snowball states are avoided.  Indeed, we should consider the possibility that the circumplanetary outer habitable edge discovered by \citet{Forgan2014a} is merely an artefact of the absence of the carbonate-silicate cycle.

Equally important to the presence and appearance of the outer habitable edge is the nature of the tidal heating.  Both \citet{Forgan_moon1} and \citet{Forgan2014a}'s climate models relied on the ``constant-phase-lag'' tidal heating prescription (see e.g. \citealt{Peale1978,Gladman1996,Greenberg2009} amongst others).  Forgan et al's previous approach fixed the tidal dissipation and rigidity parameters, ignoring the temperature dependence of the moon's structural properties, in particular its rigidity and tidal dissipation.  Typically, what we dub a ``fixed-Q'' approach underestimates the true tidal heating of the body \citep{Ross1989}, and once more this fact should give us pause when investigating the outer habitable edge.  Is the outer edge an artefact of simplistic tidal heating calculations?

We therefore present revised exomoon climate calculations using 1D latitudinal energy balance models (LEBMs), which now include the carbonate-silicate cycle and viscoelastic tidal heating \citep{Dobos2015}.  We revisit the exomoon habitable zones produced previously, in particular investigating whether the circumplanetary outer habitable edge remains a feature of exomoon habitability. 
 
In section \ref{sec:LEBM} we describe this new model setup.  Section \ref{sec:results} displays the newly derived exomoon habitable zones calculated, and sections \ref{sec:discussion} and \ref{sec:conclusions} discuss and summarise the results respectively.


\section{Latitudinal Energy Balance Modelling} \label{sec:LEBM}

\subsection{Simulation Setup}

We adopt initial conditions essentially identical to those of \citet{Forgan_moon1} and \citet{Forgan2014a}.  The star mass is $M_*=1 \msol$, the mass of the host planet $M_p = 1 \mjup$, and the mass of the moon $M_s= 1\mearth$.  This system has been demonstrated to be dynamically stable on timescales comparable to the Solar System lifetime \citep{Barnes2002}.  

The planet's orbit is given by its semi-major axis $a_p$ and eccentricity $e_p$, and the moon's orbit by $a_m$ and $e_m$ respectively.  We assume that the planet resides at the barycentre of the moon-planet system, which is satisfactory given the relatively large planet-to-moon mass ratio.  The inclination of the planet relative to the stellar equator, $i_p=0$ (i.e. the planet orbits in the $x-y$ plane).  The inclination of the moon relative to the planet's equator, i.e. the inclination of the moon relative to the $x-y$ plane, $i_m$, is zero unless stated otherwise.  The orbital longitudes of the planet and moon are defined such that $\phi_{p}=\phi_{m}=0$ corresponds to the x-axis.  We also assume that the moon's obliquity has been efficiently damped by tidal evolution, and we therefore set it to zero.

\subsection{Latitudinal Energy Balance Models with Viscoelastic Tidal Heating, Planetary Illumination and Carbonate Silicate Cycles}

\noindent The Latitudinal Energy Balance Model (LEBM) employed in this work solves the following diffusion equation:

\begin{equation} 
C \frac{\partial T}{\partial t} - \frac{\partial }{\partial x}\left(D(1-x^2)\frac{\partial T}{\partial x}\right) = (S+S_p)\left[1-A(T)\right] + \zeta - I(T), \label{eq:LEBM}
\end{equation}

\noindent where $T=T(x,t)$ is the temperature at time $t$, $x \equiv \sin \lambda$, and $\lambda$ is the latitude (between $-90\degrees$ and $90\degrees$).  This equation is evolved with the boundary condition $\frac{dT}{dx}=0$ at the poles.  The $(1-x^2)$ term is a geometric factor, arising from solving the diffusion equation in spherical geometry.

$C$ is the atmospheric heat capacity, the diffusion coefficient $D$ controls latitudinal heat redistribution, $S$ and $S_p$ are the stellar and planetary insolation respectively, $\zeta$ is the surface heating generated by tides in the moon's interior, $I$ is the atmospheric infrared cooling and $A$ is the albedo.   

Judicious selection of $D$ allows us to reproduce a fiducial Earth-Sun climate system with the correct latitudinal temperature variations measured on Earth (see e.g. \citealt{North1981, Spiegel_et_al_08}).  Planets that rotate rapidly experience inhibited latitudinal heat transport, due to Coriolis forces truncating the effects of Hadley circulation (cf \citealt{Farrell1990, Williams1997a}).  The partial pressure of $\cotwo$ also plays a role.  We follow \citet{Williams1997a} by scaling $D$ according to :

\begin{equation} 
D=5.394 \times 10^2 \left(\frac{\omega_d}{\omega_{d,\oplus}}\right)^{-2} \left(\frac{P_{\cotwo}}{P_{\cotwo,\oplus}}\right),\label{eq:D}
\end{equation}

\noindent where $\omega_d$ is the rotational angular velocity of the planet, and $\omega_{d,\oplus}$ is the rotational angular velocity of the Earth, and $P_{\cotwo,\oplus}=3.3 \times 10^{-4}  \,\rm{bar}$.  

We use a simple piecewise function to determine $P_{\cotwo}$ \citep{Spiegel2010}:

\begin{equation}
P_{\cotwo} = \begin{cases} 
10^{-2} \, \rm{bar} & T \leq 250 K \\
10^{-2 - (T-250)/27} \,\rm{bar} & 250K < T < 290K \\
P_{\cotwo, \oplus} & T \geq 290 K
\end{cases}
\end{equation}

\noindent Our prescription allows $D$ to vary with latitude, depending on the local temperature.  This is not guaranteed to produce Hadley circulation (see e.g. \citealt{Vladilo2013} for details on how $D$ can be modified to achieve this).  As we allow partial pressure of $\cotwo$ to vary, we adopt \citet{Williams1997a}'s prescription for the cooling function, $I(T,P_{\cotwo})$:

\begin{multline}
I = 9.468980 -7.714727 \times 10^{-5} \beta - 2.794778T  \\
 - 3.244753 \times 10^{-3} \beta T -3.4547406 \times 10^{-4}\beta^2 \\
 + 2.212108 \times 10^{-2} T^2 + 2.229142 \times 10^{-3} \beta^2 T  \\
 + 3.088497 \times 10^{-5} \beta T^2 - 2.789815 \times 10^{-5} \beta^2 T^2 \\
 - 3.442973 \times 10^{-3} \beta^3 - 3.361939 \times 10^{-5} T^3 \\
 + 9.173169 \times 10^{-3} \beta^3 T - 7.775195 \times 10^{-5} \beta^3 T^2 \\
 - 1.679112 \times 10^{-7} \beta T^3 + 6.590999 \times 10^{-8} \beta^2 T^3 \\  
 +1.528125 \times 10^{-7} \beta^3 T^3 - 3.367567 \times 10^{-2} \beta^4 \\
 -1.631909 \times 10^{-4} \beta^4 T + 3.663871 \times 10^{-6} \beta^4 T^2 \\
 -9.255646 \times 10^{-9} \beta^4 T^3
\end{multline}

\noindent where we have defined 

\begin{equation}
\beta = \log \left(\frac{P_{\cotwo}}{P_{\cotwo,\oplus}}\right).
\end{equation}

\noindent The diffusion equation is solved using a simple explicit forward time, centre space finite difference algorithm.  A global timestep was adopted, with constraint

\begin{equation}
\delta t < \frac{\left(\Delta x\right)^2C}{2D(1-x^2)}.  
\end{equation}

\noindent This timestep constraint ensures that the first term on the left hand side of equation (\ref{eq:LEBM})  is always larger than the second term, preventing the diffusion term from setting up unphysical temperature gradients.  The parameters are diurnally averaged, i.e. a key assumption of the model is that the moons rotate sufficiently quickly relative to their orbital period around the primary insolation source.  This is broadly true, as the star is the principal insolation source, and the moon rotates relative to the star on timescales of a few days.

The atmospheric heat capacity depends on what fraction of the moon's surface is ocean, $f_{ocean}$, what fraction is land $f_{land}=1.0-f_{ocean}$, and what fraction of the ocean is frozen $f_{ice}$:

\begin{equation} 
C = f_{land}C_{land} + f_{ocean}\left[(1-f_{ice})C_{ocean} + f_{ice} C_{ice}\right]. 
\end{equation}

\noindent The heat capacities of land, ocean and ice covered areas are 

\begin{equation} 
C_{land} = 5.25 \times 10^9  \rm{erg \, cm^{-2} K^{-1}},
\end{equation}

\begin{equation} C_{ocean} = 40.0C_{land},\end{equation}
\begin{equation} C_{ice} = \begin{cases}
9.2C_{land} &  263 K <T < 273 K \\
2C_{land} &  T<263 K \\
0.0 & T> 273 K.\\
\end{cases}. 
\end{equation}

\noindent These parameters assume a wind-mixed ocean layer of 50m \citep{Williams1997a}.  Increasing the assumed depth of this layer would increase $C_{ocean}$ (see e.g. \citealt{North1983} for details).  The albedo function is

\begin{equation} 
A(T) = 0.525 - 0.245 \tanh \left[\frac{T-268\, \mathrm K}{5\, \mathrm K} \right]. 
\end{equation}

\noindent This produces a rapid shift from low albedo ($\sim 0.3$) to high albedo ($\sim 0.75$) as the temperature drops below the freezing point of water, producing highly reflective ice sheets.  Figure 1 of \citet{Spiegel_et_al_08} demonstrates how this shift in albedo affects the potential for global energy balance, and that for planets in circular orbits, two stable climate solutions arise, one ice-free, and one ice-covered.  Spiegel et al also show that such a function is sufficient to reproduce the annual mean latitudinal temperature distribution on the Earth.

Note that we do not consider clouds in this model, which could modify both the albedo and optical depth of the system significantly.  Also, we assume that both stellar and planetary flux are governed by the same albedo, which in truth is not likely to be the case (see Discussion).

The stellar insolation flux $S$ is a function of both season and latitude.  At any instant, the bolometric flux received at a given latitude at an orbital distance $r$ is

\begin{equation}
S = q_0\cos Z \left(\frac{1 AU}{r}\right)^2,
\end{equation}

\noindent where $q_0$ is the bolometric flux received from the star at a distance of 1 AU, and $Z$ is the zenith angle:

\begin{equation} 
q_0 = 1.36\times 10^6\left(\frac{M_*}{\msol}\right)^4 \mathrm{erg \,s^{-1}\, cm^{-2}} 
\end{equation}

\begin{equation} 
\cos Z = \mu = \sin \lambda \sin \delta + \cos \lambda \cos \delta \cos h. 
\end{equation} 

\noindent $\delta$ is the solar declination, and $h$ is the solar hour angle.  As stated previously, we set the moon's obliquity $\delta_0$ to zero. The solar declination is calculated as:

\begin{equation} 
\sin \delta = -\sin \delta_0 \cos(\phi_{*m}-\phi_{peri,m}-\phi_a), 
\end{equation}

\noindent where $\phi_{*m}$ is the current orbital longitude of the moon \emph{relative to the star}, $\phi_{peri,m}$ is the longitude of periastron, and $\phi_a$ is the longitude of winter solstice, relative to the longitude of periastron.   We set $\phi_{peri,m}=\phi_a=0$ for simplicity. 

We must diurnally average the solar flux:

\begin{equation} 
S = q_0 \bar{\mu}. 
\end{equation}

\noindent This means we must first integrate $\mu$ over the sunlit part of the day, i.e. $h=[-H, +H]$, where $H$ is the radian half-day length at a given latitude.  Multiplying by the factor $H/\pi$ (as $H=\pi$ if a latitude is illuminated for a full rotation) gives the total diurnal insolation as

\begin{equation} 
S = q_0 \left(\frac{H}{\pi}\right) \bar{\mu} = \frac{q_0}{\pi} \left(H \sin \lambda \sin \delta + \cos \lambda \cos \delta \sin H\right). \label{eq:insol}
\end{equation}

\noindent The radian half day length is calculated as

\begin{equation} 
\cos H = -\tan \lambda \tan \delta. 
\end{equation}

We implement planetary illumination and eclipses of the moon in the same manner as \citet{Forgan2014a} (see also \citealt{Heller2013}).  While Heller et al allow for the planet to be in synchronous rotation and have a significant temperature difference between the dayside and nightside (expressed in the free parameter $dT_{planet}$), we assume the planet's orbit is not synchronous, and we fix $dT_{planet}=0$. The planetary albedo is fixed at 0.3.

\subsubsection{Fixed-Q and Viscoelastic Tidal Heating}

\noindent In the interest of computational expediency, we make a simple approximation for tidal heating, by firstly assuming the tidal heating per unit area is \citep{Peale1980,Scharf2006}:

\begin{equation} 
\zeta = \frac{21}{38}\frac{\rho^2_m R^{5}_m e^2_m}{\Gamma Q}\left(\frac{GM_p}{a^3_m}\right)^{5/2} 
\end{equation}

\noindent where $\Gamma$ is the moon's elastic rigidity (which we assume to be uniform throughout the body), $R_m$ is the moon's radius, $\rho_m$ is the moon's density, $M_p$ is the planet mass, $a_m$ and $e_m$ are the moon's orbital semi-major axis and eccentricity (relative to the planet), and $Q$ is the moon's tidal dissipation parameter.  

We consider both \emph{fixed-Q} tidal heating, where $Q$ and $\Gamma$ are held constant, and \emph{viscoelastic} models where $Q$ and $\Gamma$ become functions of the moon's orbital parameters.  Throughout, we assume that tidal heating occurs uniformly across the moon's surface. 

In the fixed-Q case, we assume terrestrial values: $Q=100$, $\Gamma=10^{11} \,\mathrm{dyne \, cm^{-2}}$ (appropriate for silicate rock).  In the viscoelastic case, we calculate the product  $Q\Gamma$ using the models of \citet{Dobos2015}.  In both cases, the density of the moon is fixed at $\rho_m=5 \, \mathrm{g \, cm^{-3}}$.

In the viscoelastic model the tidal heating is calculated by:

\begin{equation}
   \label{viscel}
       \dot E_\mathrm{tidal} = - \frac {21} {2} Im(k_2) \frac {R_\mathrm{m}^5 n^5 e^2} {G} \, .
\end{equation}

\noindent where $n$ is the mean motion of the moon and $Im(k_2)$ is the complex Love number, which describes structure and rheology in the satellite \citep{Segatz1988}. \citet{Henning2009} gives the value of Im($k_2$) for the Maxwell model:

\begin{equation}
   \label{Imk2}
       - Im(k_2) = \frac {57 \eta \omega} { 4 \rho g R_\mathrm{m} \left[ 1 + \left( 1 + \frac { 19 \mu } { 2 \rho g R_\mathrm{m} } \right)^2 \frac { \eta^2 \omega^2 } { \mu^2 } \right] } \, ,
\end{equation}

\noindent where $\eta$ is the viscosity, $\omega$ is the orbital frequency and $\mu$ is the shear modulus of the satellite. The temperature dependency of the viscosity and the shear modulus is described in \citet{Dobos2015}. Since only rocky bodies like Earth are considered as satellites in this work, the solidus and liquidus temperatures at which the material of the rocky body starts melting and becomes completely liquid were chosen to be 1600 K and 2000 K, respectively. We assume that disaggregation occurs at 50\% melt fraction, which leads to a breakdown temperature of 1800 K.

The viscoelastic tidal heating model also describes the convective cooling of the body. The iterative method described by \citet{Henning2009} was used for calculating the convective heat loss:

\begin{equation}
   \label{qBL}
       q_\mathrm{BL} = k_\mathrm{therm} \frac {T_\mathrm{mantle} - T_\mathrm{surf}} {\delta(T)} \, ,
\end{equation}

\noindent where $k_\mathrm{therm}$ is the thermal conductivity ($\sim 2 \mathrm{W/mK}$), $T_\mathrm{mantle}$ and $T_\mathrm{surf}$ are the temperature in the mantle and on the surface, respectively, and $\delta(T)$ is the thickness of the conductive layer. We use $\delta(T)=30 \, \mathrm{km}$ as a first approximation, and then for the iteration

\begin{equation}
   \label{delta}
       \delta(T) = \frac {d} {2 a_2} \left( \frac {Ra} {Ra_\mathrm{c}} \right)^{-1/4}
\end{equation}

\noindent is used, where $d$ is the mantle thickness ($\sim 3000$~km), $a_2$ is the flow geometry constant ($\sim 1$), $Ra_\mathrm{c}$ is the critical Rayleigh number ($\sim 1100$) and $Ra$ is the Rayleigh number which can be expressed by

\begin{equation}
   \label{Ra}
       Ra = \frac { \alpha \, g \, \rho \, d^4 \, q_\mathrm{BL} } { \eta(T) \, \kappa \, k_\mathrm{therm} } \, .
\end{equation}

\noindent Here $\alpha$ is the thermal expansivity ($\sim 10^{-4}$) and $\kappa$ is the thermal diffusivity: $\kappa = k_\mathrm{therm} / ( \rho \, C_\mathrm{p} )$ with $C_\mathrm{p} = 1260 \, \mathrm{J/(kg \, K)}$. The iteration of the convective heat flux lasts until the difference of the last two values is higher than $10^{-10} \mathrm{W/m^2}$.

We assume that with time, the tidal heating and the convective heat loss reach a stable equilibrium state. After finding the stable equilibrium temperature, the tidal heat flux is calculated, from which the $Q\Gamma$ product can be obtained.

\subsubsection{Habitability Indices \label{sec:habindex}}

\noindent We calculate habitability indices in the same manner as most groups do \citep{Spiegel_et_al_08, Vladilo2013, Forgan2014}.  The habitability function $\xi$ is:

\begin{equation} 
\xi(\lambda,t) = \left\{
\begin{array}{l l }
1 & \quad \mbox{273 K $< T(\lambda,t) <$ 373 K} \\
0 & \quad \mbox{otherwise}. \\
\end{array} \right. \end{equation}

\noindent We then average this over latitude to calculate the fraction of habitable surface at any timestep:

\begin{equation} 
\xi(t) = \frac{1}{2} \int_{-\pi/2}^{\pi/2}\xi(\lambda,t)\cos \lambda \, d\lambda. 
\end{equation}

\noindent Each simulation is allowed to evolve until it reaches a steady or quasi-steady state, and the final ten years of climate data are used to produce a time-averaged value of $\xi(t)$, $\bar{\xi}$, and the sample standard deviation, $\sigma_{\xi}$.  We use these two parameters to classify each simulations as follows:

\begin{enumerate}
\item \emph{Habitable Moons} - these moons possess a time-averaged $\bar{\xi}>0.1$, and $\sigma_{\xi} < 0.1\bar{\xi}$, i.e. the fluctuation in habitable surface is less than 10\% of the mean.
\item \emph{Hot Moons} - these moons have average temperatures above 373 K across all seasons, and are therefore conventionally uninhabitable, and $\bar{\xi} <0.1$.
\item \emph{Snowball Moons} - these moons have undergone a snowball transition to a state where the entire moon is frozen, and are therefore conventionally uninhabitable.  As with hot moons, we require $\bar{\xi}<0.1$ for the moon to be classified as a snowball, but given the nature of the snowball transition as it is modelled here, these worlds typically have $\bar{\xi}=0$.
\item \emph{Transient Moons} - these moons possess a time-averaged $\bar{\xi}>0.1$, and $\sigma_{\xi} > 0.1\bar{\xi}$, i.e. the fluctuation in habitable surface is greater than 10\% of the mean.
\end{enumerate}

\section{Results}\label{sec:results}

\subsection{The Circumstellar Exomoon Habitable Zone}

\subsubsection{The effect of the carbonate-silicate cycle}

\begin{figure*}
\begin{center}$\begin{array}{cc}
\includegraphics[scale=0.4]{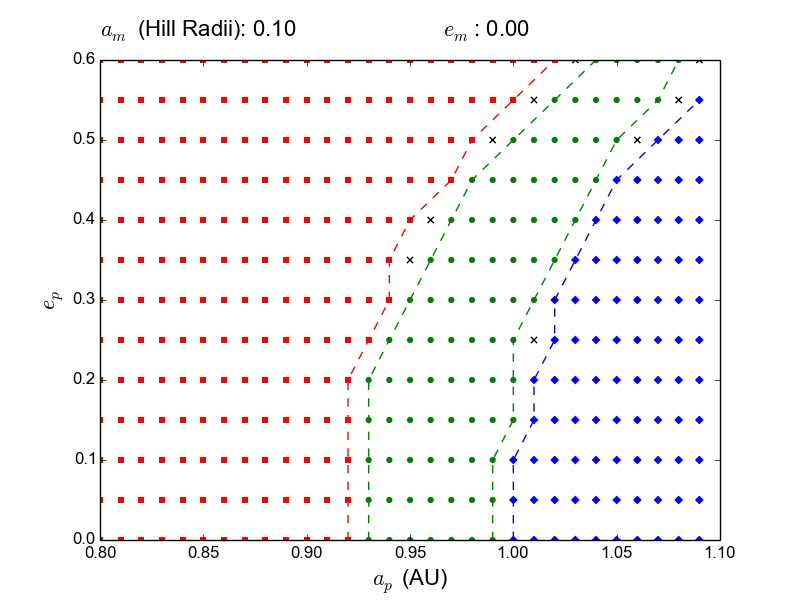} &
\includegraphics[scale=0.4]{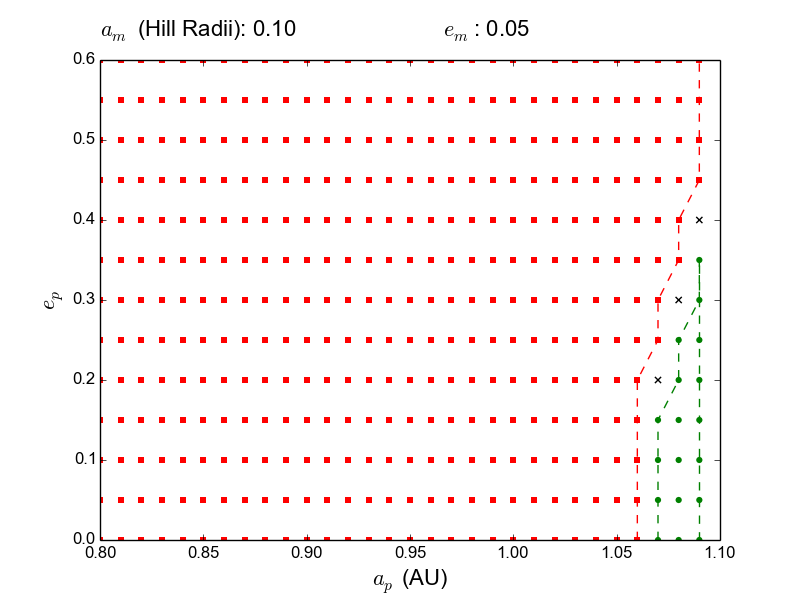} \\
\includegraphics[scale=0.4]{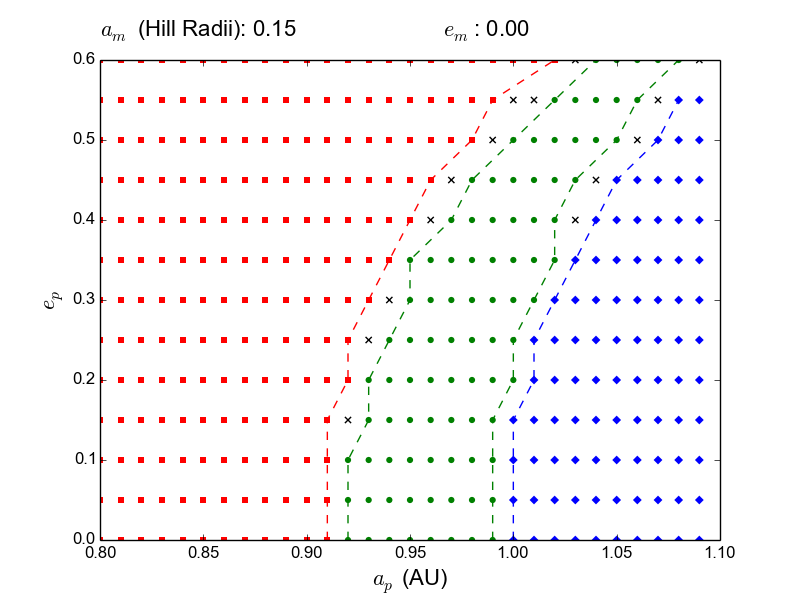} &
\includegraphics[scale=0.4]{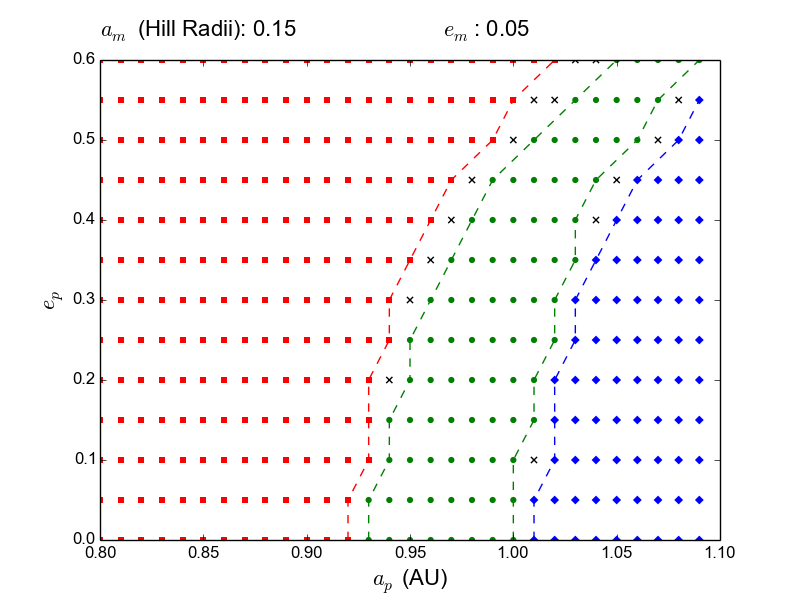} \\
\includegraphics[scale=0.4]{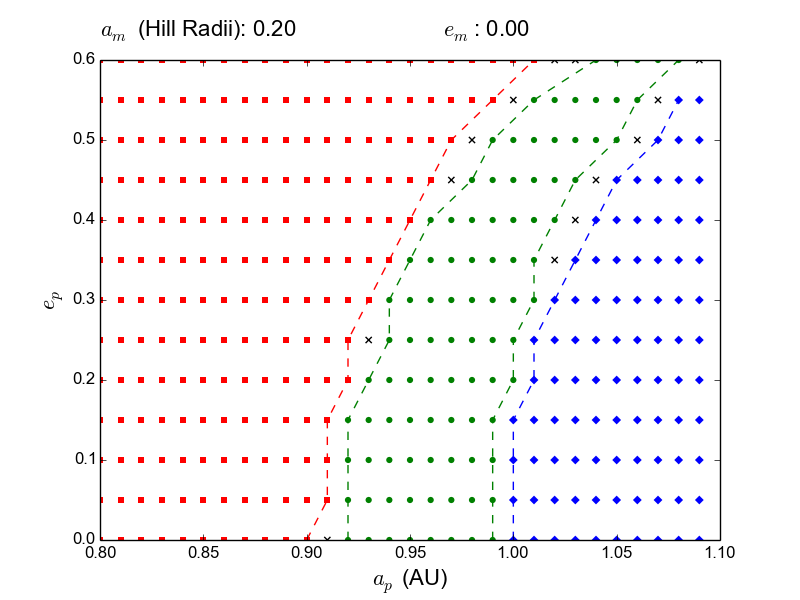} &
\includegraphics[scale=0.4]{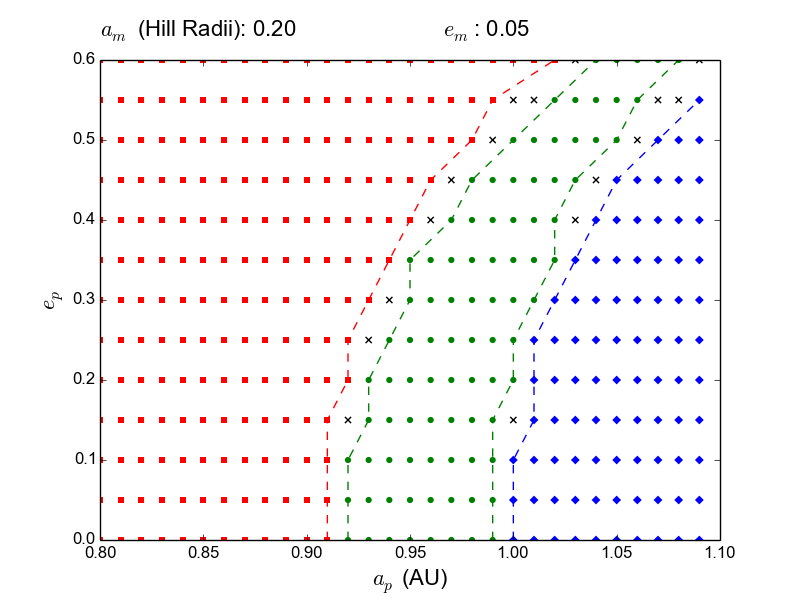} \\
\end{array}$
\caption{The exomoon habitable zone as a function of planetary parameters, with the carbonate-silicate cycle active and fixed-Q tidal heating.  The rows, from top to bottom, are moon semimajor axes of 0.1, 0.15 and 0.2 Hill Radii respectively.  The left column shows simulations where the moon eccentricity is zero, and the right column shows moons with eccentricity $e_m=0.05$.  The colours of the points indicate their classification according to the scheme in section \ref{sec:habindex}. The red squares indicate ``hot moons'' with global mean temperatures above 373 K, the blue diamonds indicate ``cold moons'', with global mean temperatures below 273 K, the green circles indicate ``habitable moons'' with mean temperatures between 273 and 373 K and low fluctuations in the mean, and the black crosses indicate ``transient moons'', where mean temperatures are in the  273-373K range, but fluctuate strongly. \label{fig:pHZ_novisc}}
\end{center}
\end{figure*}

\noindent Figure \ref{fig:pHZ_novisc} shows the classification of simulations as a function of planetary orbital parameters when the carbonate silicate cycle is active, and fixed-Q tidal heating is used.  The left column shows data for zero moon eccentricity, the right for moon eccentricity $e_m=0.05$.  

These data should be compared with Figures 2 and 3 of \citet{Forgan2014a}, which show a pronounced ``C'' shape in their outer HZ boundaries.  By contrast, our latest results possess quite a reduced sensitivity of the outer boundary to eccentricity.  While more eccentric orbits do permit larger planetary semi-major axes for habitable moons, the allowed increase in $a_p$ is not particularly large.

The inner HZ boundary has moved significantly further outward.  In our previous work without the CS cycle, the typical HZ boundary was interior of 0.85 au, whereas now this boundary exists at 0.9 au and beyond.  Replacing the \citet{Spiegel_et_al_08} cooling function with the \citet{Williams1997a} cooling function results in less efficient cooling (for a fixed $CO_2$ partial pressure).  The number of transiently habitable classifications is greatly reduced, but this is not directly due to climate modulation via the CS cycle.  This is a consequence of less efficient cooling.  Systems which would before have been able to turn away briefly from irreversible heating using the cooling function of \citet{Spiegel_et_al_08} can no longer do so, and must therefore follow trajectories towards greenhouse or habitable states.

With a small amount of moon eccentricity (right column of Figure \ref{fig:pHZ_novisc}), fixed-Q tidal heating comes into play.  The relatively inefficient cooling of the \citet{Williams1997a} function becomes increasingly evident.  At low moon semi-major axis (top row), this results in the habitable zone being pushed further outwards in planetary semimajor axis, well beyond 1 au.  Habitable moons well beyond 1 au have long been predicted, especially icy moons with subsurface liquid water oceans \citep{Reynolds1987, Scharf2006}, but this is the first indications of such behaviour in LEBM calculations of Earth-like moons at relatively low eccentricity.

As we saw in \citet{Forgan2014a}, as moon semimajor axis increases, the picture begins to look very similar to the zero moon eccentricity case, as tidal heating becomes negligible compared to the other contributors to the moon's energy budget, as can be seen in the right column of Figure \ref{fig:pHZ_novisc}.

\subsubsection{Adding viscoelastic tidal heating}

\begin{figure*}
\begin{center}$\begin{array}{c}
\includegraphics[scale=0.4]{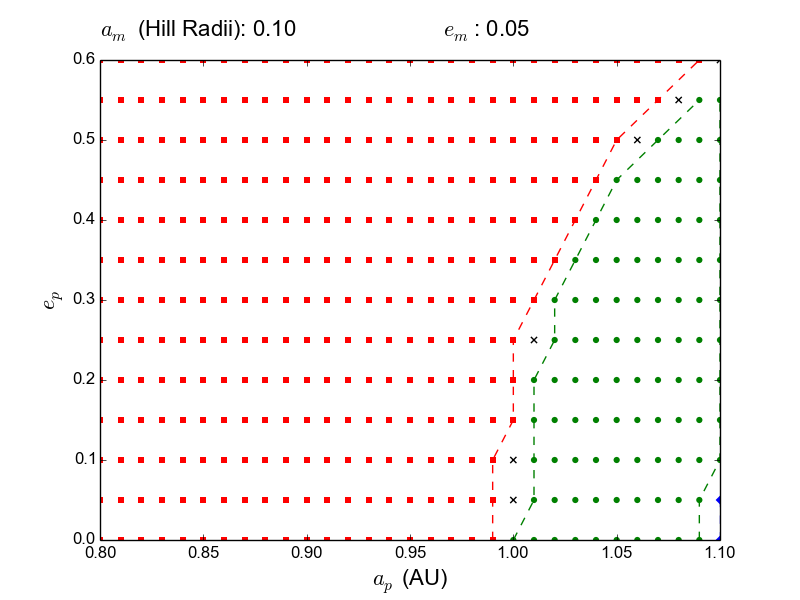} \\
\includegraphics[scale=0.4]{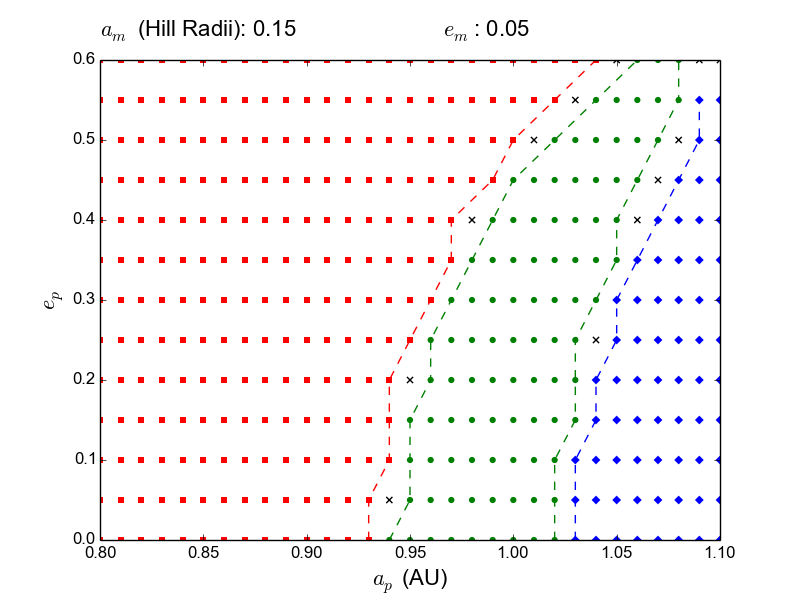}\\
\includegraphics[scale=0.4]{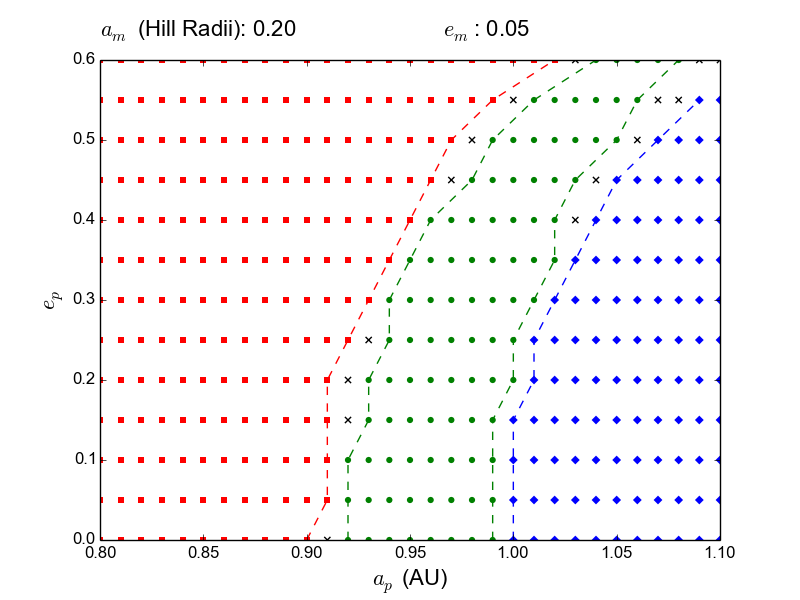} \\
\end{array}$
\caption{The effect of the carbonate silicate cycle and viscoelastic tidal heating.  This is identical to the right column of Figure \ref{fig:pHZ_novisc}, i.e. the rows from top to bottom, are moon semimajor axes of 0.1, 0.15 and 0.2 Hill Radii respectively, and the moons have eccentricity $e_m=0.05$.\label{fig:pHZ_visc}}
\end{center}
\end{figure*}

\noindent Figure \ref{fig:pHZ_visc} shows the effect of using more realistic, viscoelastic tidal heating.  We are now modelling the  inner melting of the body, which allows the values of $Q$ and $\Gamma$ to vary significantly, highlighting the fact that the values used for these quantities in the fixed-$Q$ model are somewhat arbitrary, and can vary greatly for different rocky satellites.

We can immediately see that allowing tidal parameters to vary with temperature results in a slight widening of the habitable zone.  The inner boundary once more extends as far as 0.9 au (when the moon semimajor axis is sufficiently large), again with little evidence of the ``C'' shape in the boundary curves as seen previously.  Again, low $a_m$ simulations are habitable well beyond 1au, and indeed beyond the limits of our simulation parameter space.

At large $a_m$, the differences between the fixed-$Q$ runs (bottom right of Figure \ref{fig:pHZ_novisc}) and the viscoelastic runs (bottom right of Figure \ref{fig:pHZ_visc}) become negligible.  In this limit, the tidal heating is a very small contribution to the total energy budget of the exomoon climate, and indeed both these runs are very similar to the zero eccentricity run displayed in the bottom left of Figure \ref{fig:pHZ_novisc}.

\subsection{The Exomoon Circumplanetary Habitable Zone}

\begin{figure*}
\begin{center}$\begin{array}{cc}
\includegraphics[scale=0.4]{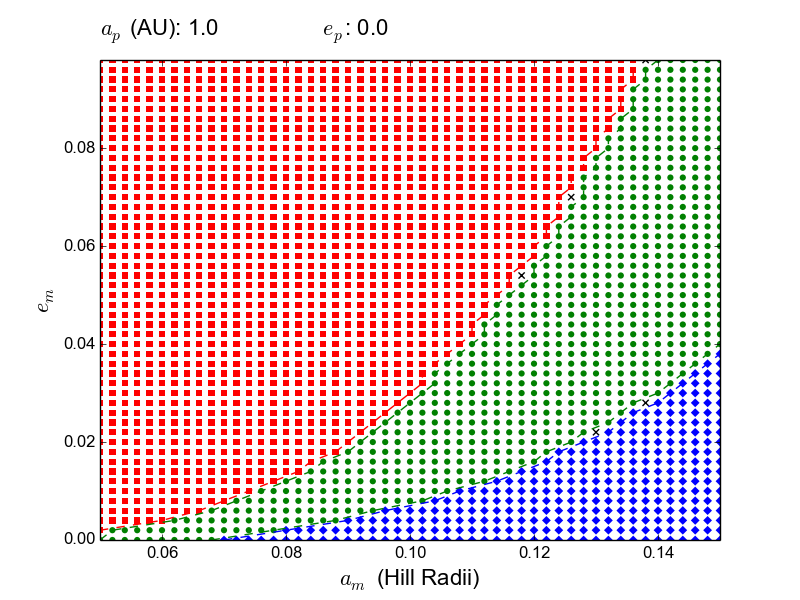} &
\includegraphics[scale=0.4]{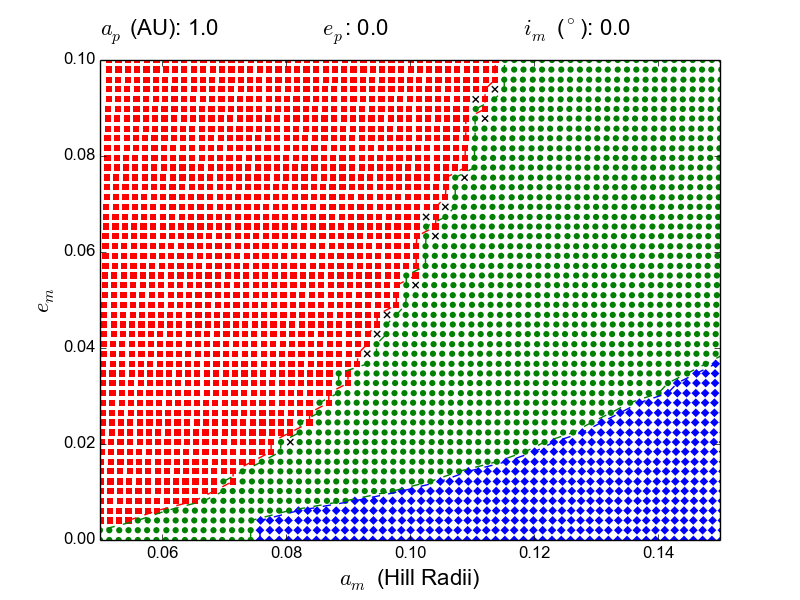} \\
\includegraphics[scale=0.4]{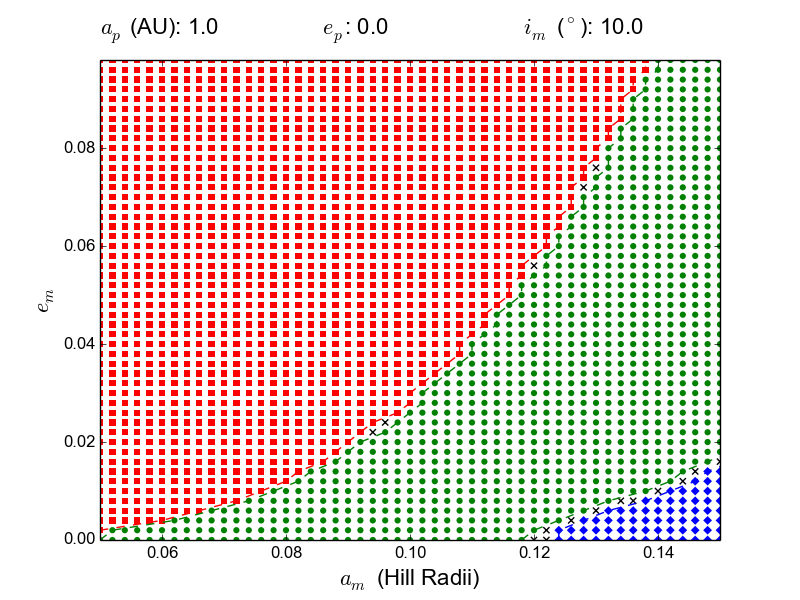} &
\includegraphics[scale=0.4]{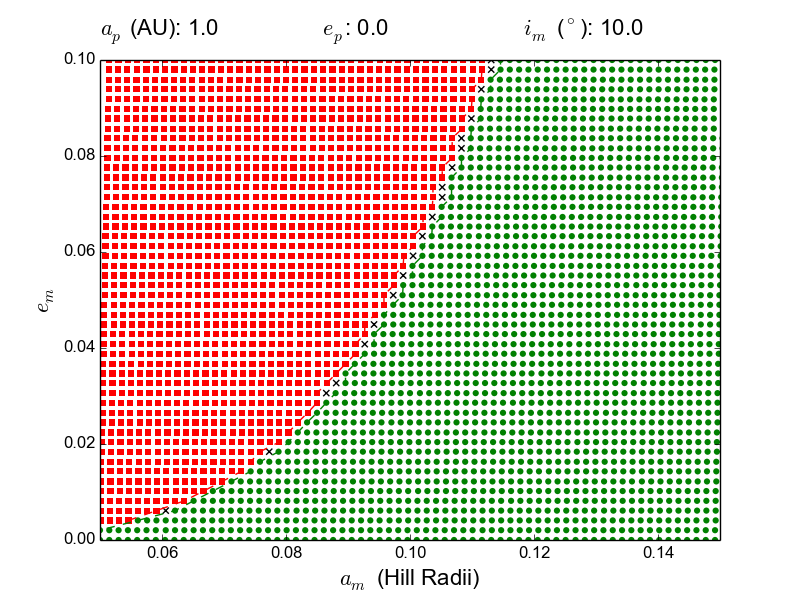}\\
\includegraphics[scale=0.4]{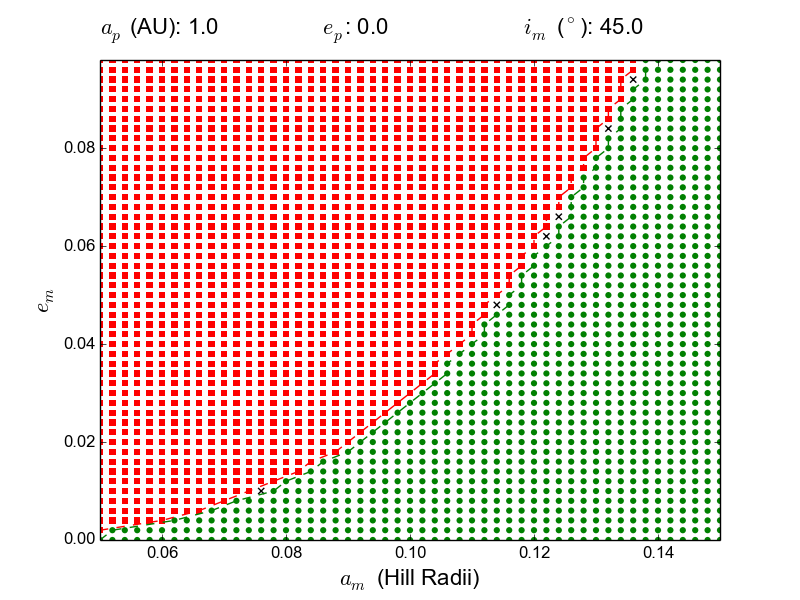} &
\includegraphics[scale=0.4]{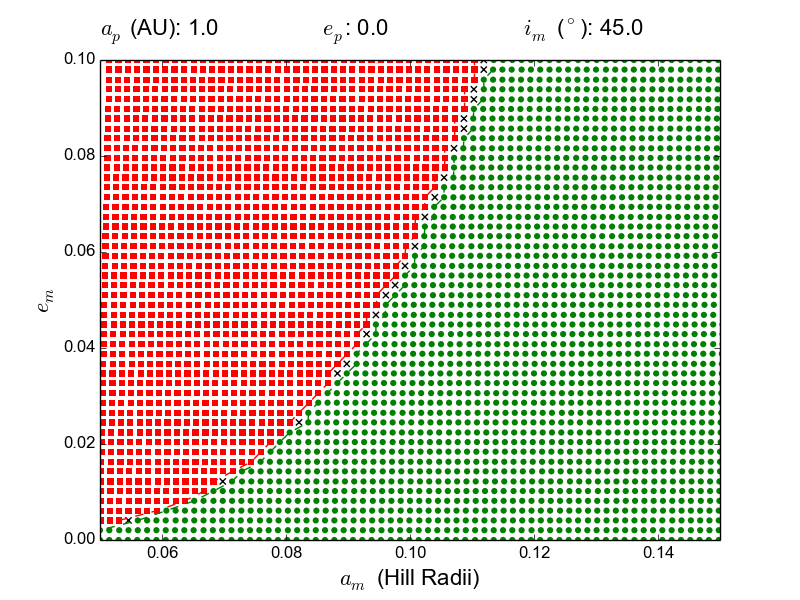}\\
\end{array}$
\caption{The habitable zone as a function of moon orbital parameters.  In all cases, the carbonate silicate cycle is active.  The left column shows the results for fixed-Q tidal heating, the right column the results for viscoelastic tidal heating.  Each row shows a different value of moon inclination: $i_m=0$ (top), $i_m = 10^\circ$ (middle) ,and $i_m=45^\circ$ (bottom). \label{fig:mHZ}}
\end{center}
\end{figure*}

\noindent We now consider the circumplanetary region, and show simulation data as a function of moon semimajor axis and eccentricity.  The planet's orbit is fixed as circular with $a_p=1$ au.  Figure \ref{fig:mHZ} shows the resulting classifications when the CS cycle is active, and when fixed-Q tidal heating is applied (left column) compared to viscoelastic tidal heating (right column).  

Beginning with the zero inclination plot for fixed-Q tidal heating (top row, left column), we can compare directly to the bottom left panel of Figure 4 in \citet{Forgan2014a}, and see immediately that the CS cycle pushes the circumplanetary habitable zone outward.  The outer boundary in previous work was calculated to be approximately 0.1 Hill Radii at $e_m=0.1$ - in the current work, the outer boundary now extends beyond 0.14 Hill Radii at $e_m=0.1$. The outer habitable edge remains at lower eccentricities, but has been pushed to higher $a_m$ along with the inner edge.  However, the width of the zone has increased from around 0.02 Hill Radii to 0.04 Hill Radii (at least, for $e_m > 0.04$).  

If viscoelastic tidal heating is used (right column of Figure \ref{fig:mHZ}), then we can see the habitable zone becoming even wider.  While the weaker tidal heating tends to pinch the habitable zone slightly at low $e_m$, at high $e_m$ this weaker tidal heating is a boon, preventing catastrophic heating of the moon.  This permits extremely wide circumplanetary HZs, at least as wide as 0.06 Hill Radii, and likely wider than we measure in this parameter space.  Perhaps most notably, for $a_m \lesssim 0.07 $ Hill Radii, the outer edge is not present at any eccentricity.  These moons experience rapid, brief eclipses which are now easily buffered by thermal inertia and the carbonate silicate cycle, regardless of the tidal heating prescription used.

If the inclination of the moon is increased (middle and bottom rows) we can see that the outer edge begins to disappear, even at relatively large moon semimajor axis. For the outer edge to become apparent, eclipses must be sufficiently frequent for their effects to be cumulative and induce a snowball state.  Increasing inclination reduces the eclipse rate (see e.g. \citealt{Heller2012}) until it is close to zero\footnote{For polar lunar orbits ($i_m=90^\circ$), the eclipse rate is not exactly zero - eclipses can still occur twice per year, when the lunar nodes align with the planet's orbital position vector}. With both implementations of tidal heating, relatively large inclinations ($45^\circ$) erase the outer habitable edge completely.  At intermediate inclinations ($10^\circ$), we see a slightly mixed picture.  The eclipse rate should be effectively zero, yet the fixed-Q tidal heating run shows that an edge still exists (albeit much further from the planet).  This is an indication that here the ice-albedo feedback mechanism is activating due to the moon's distance from the star rather than due to eclipses, which in turn may suggest that the albedo prescription we use may in fact be too simple for this scenario.

\section{Discussion}\label{sec:discussion}

\noindent While we have gone further than previous work in instituting a carbonate-silicate cycle in the model, our implementation is relatively simple.  We assume that the partial pressure of $\cotwo$ responds instantly to changes in temperature.  In practice, $\cotwo$ levels change due to imbalances between silicate weathering rates and volcanic outgassing (as well as anthropogenic sources).   \citet{Williams1997a} compute weathering rates in order to find quasi-static equilibrium values for $\cotwo$, where the relaxation timescale is assumed to be of order half a million years!  Such an approach is somewhat impractical for our aims, where we wish to run a large suite of simulations with a limited runtime needed to achieve equilibrium.  Also, we should be aware that carbonate-silicate cycle models implicitly assume a great deal about the parent body.  Models of $\cotwo$ condensation on the nightside of tidally locked planets \citep{Joshi1997,Edson2012} show the importance of a fully resolved surface, with appropriate land-ocean surfaces and interfaces.  The LEBM assumes a fixed ocean fraction at all latitudes, in effect placing limits on the extent of continents.  

Higher dimension simulations will also be of importance when attempting to calculate anisotropic tidal heating.  We have assumed that the energy produced in the interior by tidal forces is dissipated uniformly across the surface, but observations of Solar system moons such as Io show both longitudinal and latitudinal dependence in tidal heating \citep{Segatz1988}, and that this is likely to be a common feature in all tidally heated moons \citep{Peale1979,Beuthe2013}.  The precise form of this spatial distribution will depend on a) at what location in the interior dissipation occurs, and b) the transport mechanism by which this heat reaches the surface.  This becomes especially complicated for icy moons, where the ice shell around a subsurface ocean responds to tidal forcing as a membrane over a fluid layer, affecting its tidal Love number \citep{Beuthe2015}.  LEBMs could accommodate a 1D radial model of a moon's interior structure to give a latitudinal dependence on tidal heating, but are of course incapable of modelling, for example, Io's reduced tidal heating at the subplanetary point.

We should also note that our prescription for planetary illumination assumes that the moon's albedo is the same for both planetary and stellar radiation.  For example, given that the planet is likely to emit strongly in the IR, an icy moon's albedo is likely to be at least an order of magnitude lower to the planet's radiation than the star's.  A simple modification would be a ``dual passband'' system where planetary illumination calculations are carried out with a separate albedo (e.g. \citealt{Heller2013b}).  It may well be the case that the outer habitable edge can be pushed further outward by more efficient planetary illumination.

Adding the CS cycle does appear to reduce fluctuations in the moon's temperature, but we should be cognisant that the LEBM in its current form actually underestimates temperature fluctuations in general.  This is evident from studies of the Earth's Milankovitch cycles using LEBMs (\citealt{Imkeller2001,Benzi2010}, Forgan \& Mead, in prep.) which predict temperature oscillations due to Earth's orbital variation nearly an order of magnitude lower than those observed in paleoclimate data \citep{Zachos2001}.  The solution to this issue is to add short period random noise to the LEBM (to mimic weather patterns on scales less than a few days), which can allow much larger temperature variations through the phenomenon of \emph{stochastic resonance} (e.g. \citealt{Benzi1982}).  This clearly has implications for the outer circumplanetary habitable edge, which may be moved outward depending on whether such resonances can overcome the cooling effect of eclipses, once the moon moves back into sunlight.

We have seen that some exomoon orbits permit surface liquid water due to non-zero eccentricity.  However, tidal dissipation will eventually reduce the moon's eccentricity to zero.  Interactions with neighbouring moons can pump eccentricity to higher values.  Indeed, we might expect that moons experience climate cycles analogous to Earth's Milankovitch cycles, presumably on a greatly reduced timescale.  Close encounters between planets hosting satellites can also excite moon eccentricities (see e.g. \citealt{Deienno2014}).  As with planets, the habitability of an exomoon over geological timescales will depend on its dynamical landscape, and replacing fixed Keplerian orbits with an N Body calculation will provide useful insights (Forgan, submitted, Forgan \& Mead, in prep.).

\section{Conclusions }\label{sec:conclusions}

\noindent We have returned to our 1D latitudinal energy balance models of Earth-like exomoon climates, which contain stellar and planetary insolation, atmospheric circulation, infrared cooling, eclipses and tidal heating as the principal contributors to the moon's radiative energy budget.  We modify these models further by incorporating the carbonate silicate cycle, a negative feedback mechanism to regulate planetary temperatures (opposing the positive ice-albedo feedback system already present).  We also investigate improved viscoelastic models of tidal heating, which allow the moon's rigidity and tidal dissipation parameters to vary as a function of temperature.

We find that the combination of both additions results in many more potentially habitable configurations, even in our somewhat limited parameter space.  In terms of planetary orbits, the habitable zone is pushed outward, and is somewhat wider if the exomoon orbits close to the planet.  The circumplanetary habitable zone is significantly wider with viscoelastic tidal heating, and extends much further from the planet (provided the moon's eccentricity is larger than about 0.02, which is slightly less than that of Titan).  We find that there is a well-defined edge to the circumplanetary habitable zone, inside the orbital stability limit, despite the carbonate silicate cycle and viscoelastic tidal heating, due to a combination of eclipses and ice-albedo feedback.  The outer edge requires the planet and moon orbits to be quite closely aligned, so that eclipses are sufficiently frequent.  We show that if the moon's orbit is sufficiently inclined that eclipses are unlikely, the outer edge completely disappears, even for relatively large moon semimajor axis.

Given the nonlinear nature of this outer edge's origin, future work in this area must include the use of full 3D general circulation models (GCMs) with the above physics implemented, to determine the properties of this outer edge, especially in the case of slow rotating or tidally locked moons.

\section*{Acknowledgments}

DF gratefully acknowledges support from the "ECOGAL" ERC advanced grant.   VD thanks L\'aszl\'o L. Kiss for useful comments on the manuscript. VD has been supported by the Hungarian OTKA grant K104607, the Lend\"ulet-2009 Young Researchers Program of the Hungarian Academy of Sciences, the ESA PECS contract No. 4000110889/14/NL/NDe, the T\'ET-14FR-1-2015-0012 project, and the NKFIH K-115709 grant of the Hungarian National Research, Development and Innovation Office.  This work relied on the compute resources of the St Andrews MHD cluster.  The authors thank the reviewer for their insights, which helped refine and clarify the arguments made in this manuscript.

\bibliographystyle{mn2e} 
\bibliography{exomoon_cscycle}

\begin{thebibliography}{57}
\expandafter\ifx\csname natexlab\endcsname\relax\def\natexlab#1{#1}\fi

\bibitem[{Agol {et~al}\mbox{.}(2015)Agol, Jansen, Lacy, Robinson, \&
  Meadows}]{Agol2015}
Agol E., Jansen T., Lacy B., Robinson T., Meadows V., 2015, ApJ, in press

\bibitem[{Barnes \& O'Brien(2002)}]{Barnes2002}
Barnes J.~W., O'Brien D.~P., 2002, ApJ, 575, 1087

\bibitem[{Bennett {et~al}\mbox{.}(2014)Bennett, Batista, Bond, Bennett, Suzuki,
  Beaulieu, Udalski, Donatowicz, Bozza, Abe, Botzler, Freeman, Fukunaga, Fukui,
  Itow, Koshimoto, Ling, Masuda, Matsubara, Muraki, Namba, Ohnishi, Rattenbury,
  Saito, Sullivan, Sumi, Sweatman, Tristram, Tsurumi, Wada, Yock, Albrow,
  Bachelet, Brillant, Caldwell, Cassan, Cole, Corrales, Coutures, Dieters,
  {Dominis Prester}, Fouqu{\'{e}}, Greenhill, Horne, Koo, Kubas, Marquette,
  Martin, Menzies, Sahu, Wambsganss, Williams, Zub, Choi, DePoy, Dong, Gaudi,
  Gould, Han, Henderson, McGregor, Lee, Pogge, Shin, Yee, Szymański, Skowron,
  Poleski, Kozłowski, Wyrzykowski, Kubiak, Pietrukowicz, Pietrzyński,
  Soszyński, Ulaczyk, Tsapras, Street, Dominik, Bramich, Browne, Hundertmark,
  Kains, Snodgrass, Steele, Dekany, Gonzalez, Heyrovsk{\'{y}}, Kandori, Kerins,
  Lucas, Minniti, Nagayama, Rejkuba, Robin, \& Saito}]{Bennett2013}
Bennett D. {et~al.}, 2014, ApJ, 785, 155

\bibitem[{Benzi(2010)}]{Benzi2010}
Benzi R., 2010, Nonlinear Processes in Geophysics, 17, 431

\bibitem[{Benzi {et~al}\mbox{.}(1982)Benzi, Parisi, Sutera, \&
  Vulpiani}]{Benzi1982}
Benzi R., Parisi G., Sutera A., Vulpiani A., 1982, Tellus, 34, 10

\bibitem[{Beuthe(2013)}]{Beuthe2013}
Beuthe M., 2013, Icarus, 223, 308

\bibitem[{Beuthe(2015)}]{Beuthe2015}
Beuthe M., 2015, Icarus, 258, 239

\bibitem[{Canup \& Ward(2006)}]{Canup2006}
Canup R.~M., Ward W.~R., 2006, Nature, 441, 834

\bibitem[{Deienno {et~al}\mbox{.}(2014)Deienno, Nesvorn{\'{y}},
  Vokrouhlick{\'{y}}, \& Yokoyama}]{Deienno2014}
Deienno R., Nesvorn{\'{y}} D., Vokrouhlick{\'{y}} D., Yokoyama T., 2014, The
  Astronomical Journal, 148, 25

\bibitem[{Dobos \& Turner(2015)}]{Dobos2015}
Dobos V., Turner E.~L., 2015, The Astrophysical Journal, 804, 41

\bibitem[{Domingos, Winter \& Yokoyama(2006)Domingos, Winter, \&
  Yokoyama}]{Domingos2006}
Domingos R.~C., Winter O.~C., Yokoyama T., 2006, MNRAS, 373, 1227

\bibitem[{Edson {et~al}\mbox{.}(2012)Edson, Kasting, Pollard, Lee, \&
  Bannon}]{Edson2012}
Edson A.~R., Kasting J.~F., Pollard D., Lee S., Bannon P.~R., 2012,
  Astrobiology, 12, 562

\bibitem[{Farrell(1990)}]{Farrell1990}
Farrell B.~F., 1990, Journal of Atmospheric Sciences, 47, 2986

\bibitem[{Forgan(2014)}]{Forgan2014}
Forgan D., 2014, MNRAS, 437, 1352

\bibitem[{Forgan \& Kipping(2013)}]{Forgan_moon1}
Forgan D., Kipping D., 2013, MNRAS, 432, 2994

\bibitem[{Forgan \& Yotov(2014)}]{Forgan2014a}
Forgan D., Yotov V., 2014, MNRAS, 441, 3513

\bibitem[{Gladman {et~al}\mbox{.}(1996)Gladman, Quinn, Nicholson, \&
  Rand}]{Gladman1996}
Gladman B., Quinn D., Nicholson P., Rand R., 1996, Icarus, 122, 166

\bibitem[{Greenberg(2009)}]{Greenberg2009}
Greenberg R., 2009, The Astrophysical Journal, 698, L42

\bibitem[{Heller(2012)}]{Heller2012}
Heller R., 2012, A{\&}A, 545, L8

\bibitem[{Heller \& Armstrong(2014)}]{Heller2014}
Heller R., Armstrong J., 2014, Astrobiology, 14, 50

\bibitem[{Heller \& Barnes(2013)}]{Heller2013}
Heller R., Barnes R., 2013, Astrobiology, 13, 18

\bibitem[{Heller \& Barnes(2015)}]{Heller2013b}
Heller R., Barnes R., 2015, International Journal of Astrobiology, 14, 335

\bibitem[{Henning, O'Connell \& Sasselov(2009)Henning, O'Connell, \&
  Sasselov}]{Henning2009}
Henning W.~G., O'Connell R.~J., Sasselov D.~D., 2009, The Astrophysical
  Journal, 707, 1000

\bibitem[{Hippke(2015)}]{Hippke2015}
Hippke M., 2015, The Astrophysical Journal, 806, 51

\bibitem[{Iess {et~al}\mbox{.}(2014)Iess, Stevenson, Parisi, Hemingway,
  Jacobson, Lunine, Nimmo, Armstrong, Asmar, Ducci, \& Tortora}]{Iess2014}
Iess L. {et~al.}, 2014, Science, 344, 78

\bibitem[{Imkeller(2001)}]{Imkeller2001}
Imkeller P., 2001, in Stochastic Climate Models, Birkh{\"{a}}user Basel, Basel,
  pp. 213--240

\bibitem[{Joshi, Haberle \& Reynolds(1997)Joshi, Haberle, \&
  Reynolds}]{Joshi1997}
Joshi M., Haberle R., Reynolds R., 1997, Icarus, 129, 450

\bibitem[{Kipping(2009)}]{Kipping2009}
Kipping D.~M., 2009, MNRAS, 392, 181

\bibitem[{Kipping {et~al}\mbox{.}(2015)Kipping, Schmitt, Huang, Torres,
  Nesvorny, Buchhave, Hartman, \& Bakos}]{Kipping2015}
Kipping D.~M., Schmitt A.~R., Huang C.~X., Torres G., Nesvorny D., Buchhave
  L.~A., Hartman J., Bakos G.~{\'{A}}., 2015, ApJ, in press, arXiv:1503.05555

\bibitem[{Kopparapu {et~al}\mbox{.}(2013)Kopparapu, Ramirez, Kasting, Eymet,
  Robinson, Mahadevan, Terrien, Domagal-Goldman, Meadows, \&
  Deshpande}]{Kopparapu2013}
Kopparapu R.~K. {et~al.}, 2013, ApJ, 765, 131

\bibitem[{Kopparapu {et~al}\mbox{.}(2014)Kopparapu, Ramirez, SchottelKotte,
  Kasting, Domagal-Goldman, \& Eymet}]{Kopparapu2014}
Kopparapu R.~K., Ramirez R.~M., SchottelKotte J., Kasting J.~F.,
  Domagal-Goldman S., Eymet V., 2014, ApJ, 787, L29

\bibitem[{Liebig \& Wambsganss(2010)}]{Liebig2010}
Liebig C., Wambsganss J., 2010, Astronomy and Astrophysics, 520, A68

\bibitem[{Mayor \& Queloz(1995)}]{Mayor1995}
Mayor M., Queloz D., 1995, Nature, 378, 355

\bibitem[{Melosh {et~al}\mbox{.}(2004)Melosh, Ekholm, Showman, \&
  Lorenz}]{Melosh2004}
Melosh H., Ekholm A., Showman A., Lorenz R., 2004, Icarus, 168, 498

\bibitem[{Mosqueira \& Estrada(2003{\natexlab{a}})}]{Mosqueira2003a}
Mosqueira I., Estrada P.~R., 2003{\natexlab{a}}, Icarus, 163, 198

\bibitem[{Mosqueira \& Estrada(2003{\natexlab{b}})}]{Mosqueira2003}
Mosqueira I., Estrada P.~R., 2003{\natexlab{b}}, Icarus, 163, 232

\bibitem[{North, Cahalan \& Coakley(1981)North, Cahalan, \&
  Coakley}]{North1981}
North G., Cahalan R., Coakley J., 1981, Rev. Geophys. Space Phys., 19, 91

\bibitem[{North, Mengel \& Short(1983)North, Mengel, \& Short}]{North1983}
North G.~R., Mengel J.~G., Short D.~A., 1983, Journal of Geophysical Research,
  88, 6576

\bibitem[{Peale \& Cassen(1978)}]{Peale1978}
Peale S., Cassen P., 1978, Icarus, 36, 245

\bibitem[{Peale, Cassen \& Reynolds(1980)Peale, Cassen, \&
  Reynolds}]{Peale1980}
Peale S., Cassen P., Reynolds R., 1980, Icarus, 43, 65

\bibitem[{Peale, Cassen \& Reynolds(1979)Peale, Cassen, \&
  Reynolds}]{Peale1979}
Peale S.~J., Cassen P., Reynolds R.~T., 1979, Science (New York, N.Y.), 203,
  892

\bibitem[{Peters \& Turner(2013)}]{Peters2013}
Peters M.~A., Turner E.~L., 2013, ApJ, 769, 98

\bibitem[{Reynolds, McKay \& Kasting(1987)Reynolds, McKay, \&
  Kasting}]{Reynolds1987}
Reynolds R.~T., McKay C.~P., Kasting J.~F., 1987, Advances in Space Research,
  7, 125

\bibitem[{Ross \& Schubert(1989)}]{Ross1989}
Ross M., Schubert G., 1989, Icarus, 78, 90

\bibitem[{Saur {et~al}\mbox{.}(2015)Saur, Duling, Roth, Jia, Strobel, Feldman,
  Christensen, Retherford, McGrath, Musacchio, Wennmacher, Neubauer, Simon, \&
  Hartkorn}]{Saur2015}
Saur J. {et~al.}, 2015, Journal of Geophysical Research: Space Physics, 120,
  1715

\bibitem[{Scharf(2006)}]{Scharf2006}
Scharf C.~A., 2006, ApJ, 648, 1196

\bibitem[{Segatz {et~al}\mbox{.}(1988)Segatz, Spohn, Ross, \&
  Schubert}]{Segatz1988}
Segatz M., Spohn T., Ross M., Schubert G., 1988, Icarus, 75, 187

\bibitem[{Simon, Szatm{\'{a}}ry \& Szab{\'{o}}(2007)Simon, Szatm{\'{a}}ry, \&
  Szab{\'{o}}}]{Simon2007}
Simon A., Szatm{\'{a}}ry K., Szab{\'{o}} G.~M., 2007, Astronomy and
  Astrophysics, 470, 727

\bibitem[{Spiegel, Menou \& Scharf(2008)Spiegel, Menou, \&
  Scharf}]{Spiegel_et_al_08}
Spiegel D.~S., Menou K., Scharf C.~A., 2008, ApJ, 681, 1609

\bibitem[{Spiegel {et~al}\mbox{.}(2010)Spiegel, Raymond, Dressing, Scharf, \&
  Mitchell}]{Spiegel2010}
Spiegel D.~S., Raymond S.~N., Dressing C.~D., Scharf C.~A., Mitchell J.~L.,
  2010, ApJ, 721, 1308

\bibitem[{Thomas {et~al}\mbox{.}(2015)Thomas, Tajeddine, Tiscareno, Burns,
  Joseph, Loredo, Helfenstein, \& Porco}]{Thomas2015a}
Thomas P., Tajeddine R., Tiscareno M., Burns J., Joseph J., Loredo T.,
  Helfenstein P., Porco C., 2015, Icarus, 264, 37

\bibitem[{Vladilo {et~al}\mbox{.}(2013)Vladilo, Murante, Silva, Provenzale,
  Ferri, \& Ragazzini}]{Vladilo2013}
Vladilo G., Murante G., Silva L., Provenzale A., Ferri G., Ragazzini G., 2013,
  ApJ, 767, 65

\bibitem[{Ward \& Canup(2010)}]{Ward2010}
Ward W.~R., Canup R.~M., 2010, The Astronomical Journal, 140, 1168

\bibitem[{Weidner \& Horne(2010)}]{Weidner2010}
Weidner C., Horne K., 2010, A{\&}A, 521, A76

\bibitem[{Williams \& Kasting(1997)}]{Williams1997a}
Williams D., Kasting J., 1997, Icarus, 129, 254

\bibitem[{Wolszczan \& Frail(1992)}]{Wolszczan1992}
Wolszczan A., Frail D.~A., 1992, Nature, 355, 145

\bibitem[{Zachos {et~al}\mbox{.}(2001)Zachos, Pagani, Sloan, Thomas, \&
  Billups}]{Zachos2001}
Zachos J., Pagani M., Sloan L., Thomas E., Billups K., 2001, Science (New York,
  N.Y.), 292, 686

\end{thebibliography}

\appendix

\label{lastpage}

\end{document}